\documentclass[sn-mathphys,Numbered]{sn-jnl}

\usepackage{graphicx}%
\usepackage{multirow}%
\usepackage{amsmath,amssymb,amsfonts}%
\usepackage{amsthm}%
\usepackage{bm}
\usepackage{braket}
\usepackage{mathrsfs}%
\usepackage[title]{appendix}%
\usepackage{xcolor}%
\usepackage{textcomp}%
\usepackage{manyfoot}%
\usepackage{booktabs}%
\usepackage{algorithm}%
\usepackage{algorithmicx}%
\usepackage{algpseudocode}%
\usepackage{listings}%
\usepackage{soul}%
\usepackage{upgreek}
\newcommand*\diff{\mathop{}\!\mathrm{d}}

\theoremstyle{thmstyleone}%
\theoremstyle{thmstyletwo}%

\theoremstyle{thmstylethree}%

\begin{document}

\title{Zeeman polaritons as a platform for probing Dicke physics in condensed matter}


\author[1,2]{\fnm{T. Elijah} \sur{Kritzell}}
\author[1,2] {\fnm{Jacques} \sur{Doumani}}
\author[3]{\fnm{Tobias} \sur{Asano}}
\author[3]{\fnm{Sota} \sur{Yamada}}
\author[1,2]{\fnm{Fuyang} \sur{Tay}}
\author[4] {\fnm{Hongjing} \sur{Xu}}
\author[4,5] {\fnm{Han} \sur{Yan}}
\author[3,6]{\fnm{Ikufumi} \sur{Katayama}}
\author[3,6]{\fnm{Jun} \sur{Takeda}}
\author[4,5]{\fnm{Andriy} \sur{Nevidomskyy}}
\author[7]{\fnm{Hiroyuki} \sur{Nojiri}}
\author[3,6]{\fnm{Motoaki} \sur{Bamba}}
\author*[2,5,8]{\fnm{Andrey} \sur{Baydin}}\email{baydin@rice.edu}
\author*[2,4,5,8,9]{\fnm{Junichiro} \sur{Kono}}\email{kono@rice.edu}

\affil[1]{\orgdiv{Applied Physics Graduate Program, Smalley-Curl Institute}, \orgname{Rice University}, \orgaddress{\street{6100 Main Street}, \city{Houston}, \postcode{77005}, \state{TX}, \country{USA}}}

\affil[2]{\orgdiv{Department of Electrical and Computer Engineering}, \orgname{Rice University}, \orgaddress{\street{6100 Main Street}, \city{Houston}, \postcode{77005}, \state{TX}, \country{USA}}}

\affil[3]{\orgdiv{Department of Physics}, \orgname{Yokohama National University}, \orgaddress{\street{79-5 Tokiwadai, Hodogaya-ku}, \city{Yokohama}, \postcode{240-8501}, \state{Kanagawa}, \country{Japan}}}

\affil[4]{\orgdiv{Department of Physics and Astronomy}, \orgname{Rice University}, \orgaddress{\street{6100 Main Street}, \city{Houston}, \postcode{77005}, \state{TX}, \country{USA}}}

\affil[5]{\orgdiv{Smalley-Curl Institute}, \orgname{Rice University}, \orgaddress{\street{6100 Main Street}, \city{Houston}, \postcode{77005}, \state{TX}, \country{USA}}}

\affil[6]{\orgdiv{Institute for Multidisciplinary Sciences}, \orgname{Yokohama National University}, \orgaddress{\street{79-5 Tokiwadai, Hodogaya-ku}, \city{Yokohama}, \postcode{240-8501}, \state{Kanagawa}, \country{Japan}}}

\affil[7]{\orgdiv{Institute for Materials Research}, \orgname{Tohoku University}, \orgaddress{\street{2-1-1 Katahira, Aoba-ku}, \city{Sendai}, \postcode{980-8577}, \state{Miyagi}, \country{Japan}}}

\affil[8]{\orgdiv{Rice Advanced Materials Institute}, \orgname{Rice University}, \orgaddress{\street{6100 Main Street}, \city{Houston}, \postcode{77005}, \state{TX}, \country{USA}}}

\affil[9]{\orgdiv{Department of Materials Science and NanoEngineering}, \orgname{Rice University}, \orgaddress{\street{6100 Main Street}, \city{Houston}, \postcode{77005}, \state{TX}, \country{USA}}}

\abstract{
The interaction of an ensemble of two-level atoms and a quantized electromagnetic field, described by the Dicke Hamiltonian, is an extensively studied problem in quantum optics. However, experimental efforts to explore similar physics in condensed matter typically employ bosonic matter modes (e.g., phonons, magnons, and plasmons) that are describable as simple harmonic oscillators, i.e., an infinite ladder of equally spaced energy levels. Here, we examine ultrastrong coupling between a coherent light mode and an ensemble of paramagnetic spins, a finite-multilevel system, in Gd$_3$Ga$_5$O$_{12}$. The electron paramagnetic resonance of Gd$^{3+}$ ions is tuned by a magnetic field into resonance with a Fabry--P\'erot cavity mode, resulting in the formation of spin--photon hybrid states, or Zeeman polaritons. We observe that the light--matter coupling strength, measured through the vacuum Rabi splitting, decreases with increasing temperature, which can be explained by the temperature-dependent population difference between the lower and higher-energy states, a trait of a finite-level system. This finding demonstrates that a spin--boson system is more compatible with the Dicke model and has advantages over boson--boson systems for pursuing experimental realizations of phenomena predicted for ultrastrongly coupled light--matter hybrids.
}

\maketitle

In the last decade, ultrastrong light--matter coupling~\cite{Forn2019RevMod, Kockum2019NatRev} has stimulated much interest among many experimental~\cite{Gao2018NatPhot, Li2018NatPhot, Zhang2016NatPhys, Jarc2023Nature} and theoretical~\cite{Schlawin2022Applied, GarciaVidal2021Science, Curtis2019PRL, Sentef2020PRR} disciplines, owing to the exciting potentials in quantum information science~\cite{Li2022PhotInsight, WangJAPP2020, LachanceApex2019, HarderBook}, cavity quantum electrodynamics (QED)~\cite{Garraway2011Philos, Forn2019RevMod, Kockum2019NatRev}, and condensed matter physics~\cite{Schlawin2022Applied, GarciaVidal2021Science}. At the heart of quantum-mechanical photon--atom interactions lies the Dicke model~\cite{Dicke1954PhysRev}, which describes the cooperative interaction of an ensemble of two-level atoms with a single mode of light. Among the most prominent implications of the model is a second-order phase transition known as the superradiant phase transition~\cite{Hepp1973AOP, Wang1973PRA}, which occurs when the coupling strength reaches a threshold value that is comparable to the bare frequencies of light and matter, i.e., the ultrastrong coupling (USC) regime~\cite{Forn2019RevMod, Kockum2019NatRev}.

Attempts to experimentally realize the superradiant phase transition in a cavity-condensed matter system are limited by the no-go theorem stemming from the presence of the diamagnetic (or the $A^2$) term in the electric-dipole-based light--matter interaction Hamiltonian~\cite{RzazewskiPRL1975, Bialynicki1979PRA}. 

However, a no-go theorem has yet to be applied to a magnetic-dipole-based light--matter hybrid system~\cite{Kim2024Arxiv, Bamba2022CommPhys}. To date, strong light--matter coupling has been observed using magnon resonances in ferromagnetic~\cite{Huebl2013PRL, TabuchiEtAl2014PRL, ZhangEtAl2014PRL} and antiferromagnetic~\cite{Li2018Science, Grishunin2018ACSPhot, BaydinEtAl2023PRR, Bialek2023PRA, kritzell2023AOM} materials. However, much like other typical cavity-coupled many-body condensed matter systems, these deviate in character from the system originally described by Dicke~\cite{Dicke1954PhysRev} in the sense that they involve bosonic matter modes (i.e., magnons)~\cite{Li2018NatPhot, Zhang2016NatPhys, Huebl2013PRL, Scalari2012Science, Roh2023NanoLetters, Bourcin2023PRB, Berkmann2024ACSPhot}. Such a matter mode is prototypically modeled as a simple harmonic oscillator, which, quantum mechanically, is an infinite ladder of equally spaced energy levels. Therefore, the light--matter coupling is described as a boson-boson interaction through the Hopfield Hamiltonian~\cite{Hopfield}; see the right panel of Figure\,\ref{Fig:ExpScheme}\textbf{a}. 

The Dicke model, in comparison, describes an ensemble of two-level atoms resonantly coupled with photons in a single-mode cavity; see the left panel of Figure\,\ref{Fig:ExpScheme}\textbf{a}. For example, because the number of energy levels on the matter side is finite, the coupling strength is expected to change with temperature. When the temperature increases, the vacuum Rabi splitting (VRS) decreases as the coupling tends to vanish due to the saturation of the steady-state population difference between the lower and upper levels through thermal excitation~\cite{yuenzhou2023Arxiv, Cwik2016PRA, Deng2002Science}. Conversely, in a simple harmonic oscillator, the population can never saturate, and thus, the coupling strength remains independent of the temperature.

Electron paramagnetic resonance (EPR) is an excellent and, to date, an unexplored avenue to examine strong light--condensed matter coupling. In a paramagnetic material in an applied magnetic field, the orientation of an unpaired spin can be changed via the absorption of a photon whose energy is equal to the magnetic-field-induced spin Zeeman splitting. When the coupling strength exceeds the matter and cavity decay rates, the system enters the strong coupling regime, leading to the formation of spin--photon hybrid states, which can be referred to as Zeeman polaritons, exhibiting a finite VRS in the frequency domain. The finite number of possible spin orientations ensures that the EPR excitation retains the multilevel character while the magnetic-dipole nature of EPR retains the advantage of being unaffected by the $A^2$ term in the interaction Hamiltonian. 

Here, we studied spin--boson USC using an ensemble of paramagnetic spins in Gd$_3$Ga$_5$O$_{12}$ (or gadolinium gallium garnet, GGG). We tuned the frequency of the EPR of Gd$^{3+}$ ions by a magnetic field to resonate with a Fabry--P\'erot (FP) cavity mode, which resulted in the appearance of Zeeman polaritons with clear VRS. The value of the VRS was a function of temperature, monotonically decreasing with increasing temperature. At low temperatures, USC was achieved with VRS of $\sim$60\% of the resonance frequency. We explain the observed temperature dependence as a consequence of the matter having a finite number of levels. Namely, the population difference between the lower- and higher-energy states decreases with increasing temperature, reaching zero at infinite temperature, which is a characteristic of a finite-level system in thermal equilibrium. These results demonstrate that a spin--boson system, which is more compatible with the Dicke model, has significant advantages over boson--boson systems in realizing phenomena predicted for ultrastrongly coupled light--matter hybrids~\cite{Forn2019RevMod, Kockum2019NatRev}.

\section*{Experimental Scheme}

\begin{figure}[!htb]
     \centering
     \includegraphics[width=1\linewidth]{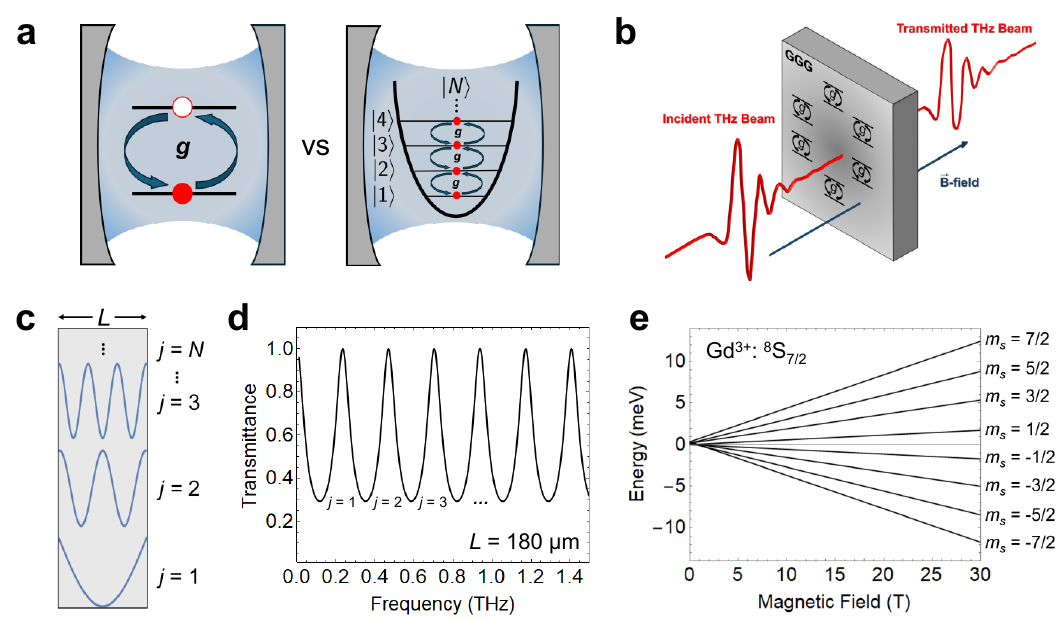}
     \caption{\textbf{a},~Typically, the matter resonance in studies of cavity--condensed matter interactions can be modeled by a simple harmonic oscillator (shown right), describable by the Hopfield Hamiltonian. We utilize a finite-multilevel system -- an ensemble of paramagnetic spins -- closer to the original atom--photon system described by the Dicke model (left). \textbf{b},~Schematic diagram of the experiment. The cavity magnetic field interacts with the ensemble of spins within the sample. The cavity dynamics are probed with a weak THz pulse in Faraday geometry. \textbf{c},~Schematic diagram of a Fabry-P{\'{e}}rot cavity formed by the GGG crystal itself with an appropriate thickness. Multiple reflections between the front and back sample--air interfaces lead to a resonant cavity magnetic field inside the sample. \textbf{d},~Transmittance spectra for a Fabry-P{\'{e}}rot cavity. Each transmission maximum in the spectrum corresponds to a cavity mode. \textbf{e},~The energies and corresponding spin configurations of the lowest eight energy levels of Gd$^{3+}$ ions inside the GGG crystal as a function of the magnetic field.}\label{Fig:ExpScheme}
\end{figure}

GGG is an isotropic magnet, with magnetic Gd$^{3+}$ ions remaining paramagnetic at temperatures above 1.5\,K~\cite{Bedyukh1999LTPhys}. Figure\,\ref{Fig:ExpScheme}\textbf{e} shows the energies of the energy levels of Gd$^{3+}$ ions in the lowest-energy manifold inside the GGG crystal as a function of magnetic field. The spin configuration corresponding to each energy level is depicted on the right side of the figure. In the ground state of four Kramers doublets, there are seven allowed energy transitions corresponding to $\Delta m_\mathrm{s} = 1$, where $m_\mathrm{s}$ is the value of the total spin~\cite{Deen2015PRB}. These transitions correspond to one triply degenerate and two doubly degenerate EPR modes due to crystal fields, with Zeeman splitting frequencies, $\omega_\text{EPR}$, on the order of 0.1 terahertz (THz) at high magnetic fields above 5\,T~\cite{Dai1988PRD}.

To form a FP cavity, rather than using mirrors to form a cavity, we utilize the sample--vacuum interfaces~\cite{BialekEtAl2020PRB, kritzell2023AOM}. The sample was polished so that the thickness became comparable to the wavelength of the light mode; see Figure\,\ref{Fig:ExpScheme}\textbf{c}. We obtained an array of equally spaced cavity modes, with frequency and linewidth related to the refractive index, $n \approx 3.8$, and thickness, $L$, of the cavity material such that the cavity mode frequencies $\omega_\text{cav}^{(j)} / (2 \pi) = j c/(2 n L)$, where $j$ is the index of the mode and $c$ is the speed of light, and the linewidth $\delta\omega_\text{cav} = \omega_{\text{cav}}\sqrt{r} / (1-r)$, where $r$ is the reflection coefficient of the surface, $r = |\frac{n - 1}{n + 1}|^4$. Each mode corresponds to a harmonic of the electric field profile within the cavity; see Figure\,\ref{Fig:ExpScheme}\textbf{d}. While the Dicke model does not require the cavity and EPR modes to be resonant to achieve novel phenomena, we focus on the resonant condition so that the coupling strength can be experimentally characterized.

We utilized two samples with different thicknesses -- Samples 1 and 2 -- so that the most relevant cavity mode frequencies are compatible with the spectral and magnetic field ranges available in two THz magnetospectroscopy setups; see Methods. We thinned down the crystals to $L$ = 180\,$\upmu$m (Sample 1) and 129\,$\upmu$m (sample 2), respectively, using a wafer polisher with coarse-grit silicon carbide paper. Doing so yielded FP cavity modes in Sample 1 (Sample 2) with a mode spacing $\Delta\omega_\text{cav}/(2\pi) = \omega_\text{cav}^{(j+1)}/(2\pi) - \omega_\text{cav}^{(j)}/(2\pi)= c/(2nL)$ of 218\,GHz (304\,GHz) and a full-width-at-half-maximum $\delta\omega_\text{cav}$ of 103\,GHz (160\,GHz). 
In Sample 1, we tuned $\omega_\text{EPR}$ by a magnetic field to resonate with $\omega_\text{cav}^{(1)}$ while in Sample 2 $\omega_\text{EPR}$ was tuned to be resonant with $\omega_\text{cav}^{(2)}$.

\section*{Observation of Zeeman Polaritons in the Ultrastrong Coupling Regime} \label{Sec:data}

\begin{figure}[H]
     \centering
     \includegraphics[width=1\textwidth]{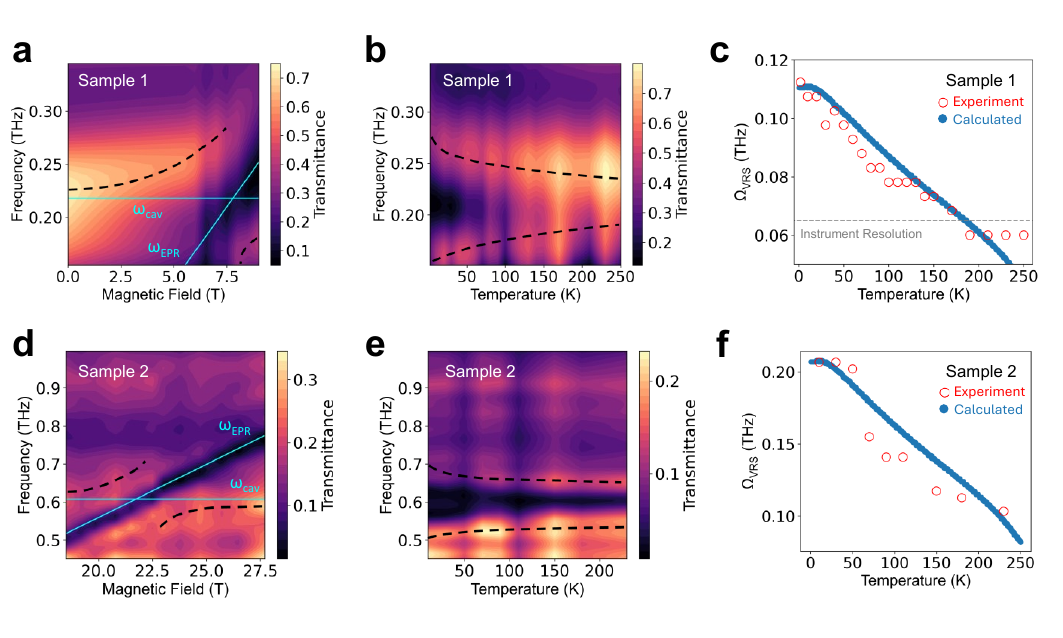}
     \caption{
     \textbf{a},~Transmittance spectra for Sample 1 measured at various magnetic fields up to 9\,T at 1.5\,K. Anticrossing behavior indicating the formation of Zeeman polaritons is revealed. Zero-detuning between the cavity and EPR modes occurs at 7.8\,T, with a frequency of 218\,GHz. The dashed lines following the lower and upper polariton branches are guides to the eye. The solid line indicates the bare EPR and FP mode frequencies. \textbf{b},~Transmission spectra at various temperatures at the zero-detuning point. The upper and lower polaritons diverge at low temperatures, with a maximum vacuum Rabi splitting of $\Omega_\text{VRS}$ = 112\,GHz. The dashed lines following the upper and lower polariton branches are guides to the eye. \textbf{c},~Experimental (red points) and calculated (blue dashed line) vacuum Rabi splitting as a function of temperature for the $j=1$ mode of Sample 1. The gray dashed line denotes the spectrometer resolution limit. Error bars for the experimental points are inside the markers. \textbf{d},~Transmission spectra for Sample 2 measured at various magnetic fields up to 30\,T at 235\,K. Anticrossing behavior indicating the formation of Zeeman polaritons is revealed. Zero-detuning between the FP cavity and EPR modes occurs at 21.5\,T, with a frequency of 608\,GHz, \textbf{e},~Transmission spectra at various temperatures at the zero-detuning point. The maximum Rabi splitting is $\Omega_\text{VRS}$ = 206\,GHz. \textbf{f},~Vacuum Rabi splitting as a function of temperature for the $j=2$ mode of Sample 2.
     } \label{Fig:Data2}
\end{figure}

Figure~\ref{Fig:Data2}\textbf{a}~depicts transmittance spectra for Sample 1, measured at various magnetic fields up to 9\,T at 1.5\,K. We tuned the matter mode by applying a magnetic field perpendicular to the surface plane of the crystal. At 7.8\,T, the frequencies of the fundamental cavity mode and the EPR mode coincided, i.e., $\omega_\text{cav}^{(1)}/(2\pi) = \omega_\text{EPR}/(2\pi) = 218$\,GHz, revealing an anticrossing, which indicates the formation of Zeeman polaritons. The magnitude of the observed VRS, $\Omega_\text{VRS}$, is 112\,GHz. Fitting the temperature dependence of the VRS to the model described in 
Section ``Theory of Zeeman Polaritons,''
we obtain a zero-temperature coupling strength of $g_0 / (2\pi) = 47.5$\,GHz, 
corresponding to a normalized coupling strength of $\eta \equiv g_0 / \omega_\text{cav} = 0.22$. The observed VRS is notably larger than $\Omega_{\text{VRS}} \equiv 2 g_0$ expected in the limit of $\eta \ll 1$. This discrepancy is likely due to the ultrastrong coupling $\eta > 0.1$, poor separation of cavity harmonics, as well as large cavity loss.

Figure~\ref{Fig:Data2}\textbf{b}~depicts transmittance spectra for Sample 1, measured at various temperatures from 1.5\,K to 250\,K. Holding the magnetic field constant at 7.8\,T, where $\omega_\text{cav}^{(1)}/(2\pi) = \omega_\text{EPR}/(2\pi) = 218$\,GHz, and varying the temperature of the sample, we find that the VRS decreases with increasing temperature. As the temperature is raised, the upper and lower polariton branches begin to merge at room temperature. This behavior contrasts that seen in other condensed matter cavity QED systems~\cite{Zhang2016NatPhys, Berkmann2024ACSPhot, Golovchanskiy2021SciAdv} in the USC regime, where the VRS remains constant as a function of temperature because of the simple harmonic oscillator nature of the matter excitation.

The red markers in Figure~\ref{Fig:Data2}\textbf{c}~depict the VRS value as a function of temperature for the case illustrated in Figure~\ref{Fig:Data2}\textbf{b}. At low temperatures, the VRS reaches a maximum value of 112\,GHz ($\eta = 0.22$). As the temperature increases, the VRS decreases. Above 190\,K, the VRS is comparable to or less than the spectrometer resolution of 65\,GHz and cannot be further estimated.
Figure~\ref{Fig:Data2}\textbf{d}~depicts transmittance spectra for Sample 2, measured at various magnetic fields up to 25\,T at 235\,K. We tuned the matter mode by applying a pulsed magnetic field~\cite{Tay2022JPSJ, Noe2013RevSciInst, Noe2016OptExp} perpendicular to the surface plane of the crystal. Similar to Fig.~\ref{Fig:Data2}\textbf{a}, Zeeman polaritons are formed at 21.5\,T, where an anticrossing is revealed for $\omega_\text{cav}^{(2)} /(2\pi) = \omega_\text{EPR}/(2\pi) =$ 608\,GHz. The value of the observed VRS is 103\,GHz. Fitting the VRS, we obtain $g_0/(2\pi) = 79.3$\,Ghz, corresponding to $\eta = 0.13$.

Figure~\ref{Fig:Data2}\textbf{e}~depicts transmittance spectra for Sample 2, measured at various temperatures from 12\,K to 235\,K.  Holding the magnetic field constant at 21.5\,T, where $\omega_\text{cav}^{(2)}/(2\pi) = \omega_\text{EPR}/(2\pi) =$ 608\,GHz, and varying the temperature of the sample, we find that the VRS decreases with increasing temperature. 
The red markers in Figure~\ref{Fig:Data2}\textbf{f}~depicts the VRS value as a function of temperature for the case illustrated in Figure~\ref{Fig:Data2}\textbf{e}. At low temperatures, the VRS reaches a maximum value of 207\,GHz ($\eta$ = 0.13). As the temperature increases, the VRS decreases. However, in this case, the VRS value decreases more slowly than in the case of Sample 1 (Figure \ref{Fig:Data2}\textbf{c}). The value of the normalized coupling strength $\eta$ achieved is lower than that obtained for the fundamental mode at low temperature in Sample 1 but holds at higher values as the temperature is raised.

\section*{Theory of Zeeman Polaritons}\label{Sec:g0}

\begin{figure}[!htb]
     \centering
     \includegraphics[width=1\textwidth]{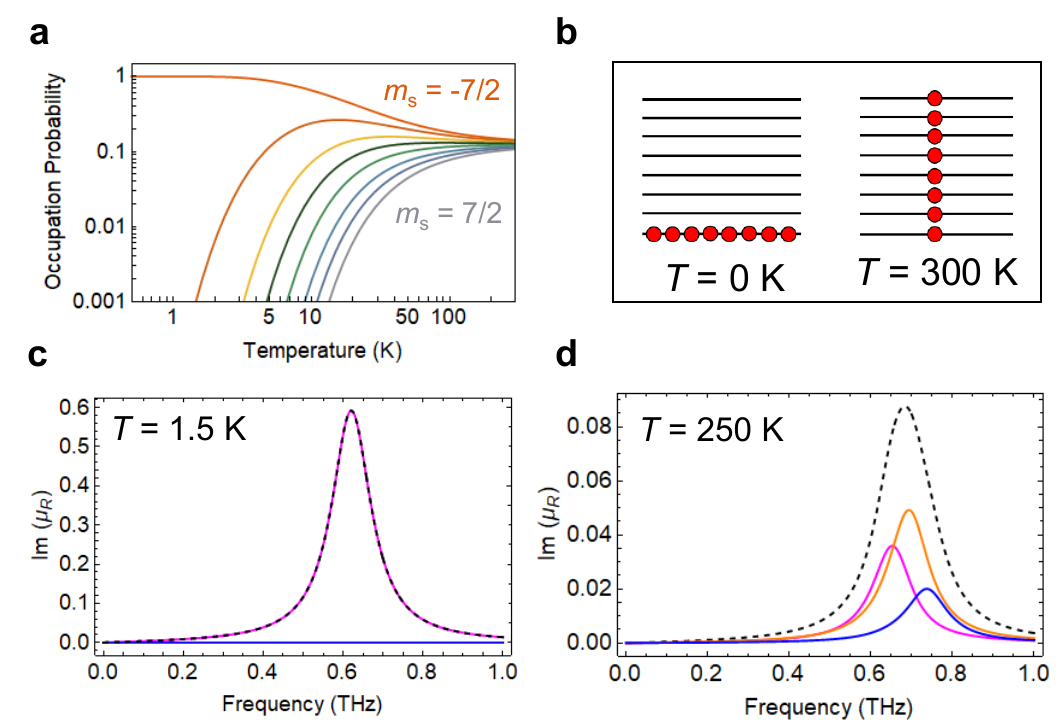}
     \caption{
     \textbf{a}, Probability of each state as a function of temperature. In the zero-temperature limit, each ion will remain in the $m_\text{s} = -7/2$ energy level, corresponding to the highest coupling strength. At high temperatures, the energy levels will saturate with equal distribution (illustrated in \textbf{b}), and the coupling will disappear. \textbf{c}, At $T$ = 1.5\,K, the imaginary permeability shows a packet dominated by the lowest-energy absorption, corresponding to the $m_\text{s} = -7/2 \rightarrow -5/2$ transition. \textbf{d}, At $T$ = 250\,K, the imaginary permeability shows a packet consisting of all three degenerate Zeeman transitions. As the temperature increases, the packet gains contributions from all level transitions. 
     }\label{Fig:Theory}
\end{figure}

In the presence of an external DC magnetic flux density $B^{\text{DC}}$, the degeneracy of $\mathrm{Gd}^{3+}$ spin levels in GGG is lifted, as shown in Figure\,\ref{Fig:ExpScheme}\textbf{e}~\cite{Dai1988PRD}. The peak positions of the Zeeman polaritons are governed by GGG's relative permeability $\mu_\mathrm{r}(\omega)=1/(1-\chi(\omega))$ or magnetic susceptibility $\chi(\omega)$ relating the magnetization $M_x(\omega)=\chi(\omega)B_x(\omega)/\mu_0$ to the AC magnetic flux density $B_x(\omega)$ perpendicular to $B^{\text{DC}} \parallel z$ in our experimental setup. We derived $\chi(\omega)$ by linear response theory~\cite{Kubo1957-nz} with the Zeeman interaction Hamiltonian
\begin{align} \label{eq:Zeeman}
    \hat{\mathcal{H}}_{\text{I}} = -\hat{d_x}B_x
\end{align}
describing the coupling between the magnetic moment $\hat{d}_x=-\mathfrak{g}_{\text{e}} \mu_\text{B} \hat{s}_x/ \hbar$ of a spin-$7/2$ particle and the AC magnetic flux density $B_x$, while the unperturbed Hamiltonian gives the energy levels in Figure\,\ref{Fig:ExpScheme}\textbf{e} including the Zeeman interaction with $B^{\text{DC}}$. The magnetic susceptibility is derived as (see details in Methods)
\begin{align} \label{eq:chi}
   \chi(\omega)
&= \sum_{m_\text{s}=-s+1}^s \frac{4g_{m_\text{s}}{}^2 }{\left( \frac{E_{m_\text{s}} - E_{m_\text{s}-1}}{\hbar} \right)^2 - \omega^2 - i \gamma \omega}.
\end{align}
In this way, $\chi(\omega)$ is represented by a sum of Lorentzians associated with all possible spin $s=7/2$ configurations, which depend on the energy differences between the Zeeman levels associated with energy $E_{m_\text{s}}$ as well as the matter damping rate $\gamma$.
Because the population of each energy level varies with the temperature $T$ (Figure~\ref{Fig:Theory}\textbf{a}), the coupling strength for the $m_\text{s}$ transition
\begin{align} \label{eq:g2}
    g_{m_\text{s}} = g_0 \sqrt{\frac{(s+m_\text{s})(s-m_\text{s}+1)}{7} \frac{E_{m_\text{s}} - E_{m_\text{s}-1}}{\hbar \omega_\text{EPR}} (P_{m_\text{s}-1} - P_{m_\text{s}})}
\end{align}
is a function of $B^{\text{DC}}$ and $T$ through $E_{m_\text{s}}$ and the occupation probability $P_{m_\text{s}}=\text{exp} \{ -E_{m_\text{s}} / (k_\text{B}T)\}/Z$, where $k_\text{B}$ is the Boltzman constant and $Z$ is the partition function.
Here, $\omega_\text{EPR}$ (determined by $B^{\text{DC}}$) is the EPR frequency associated with the transition between the two lowest Zeeman levels.
As $T \rightarrow 0$, $g_{-5/2}$ approaches the zero-temperature coupling strength
\begin{align} \label{Eq:g0}
        g_0 = \frac{1}{2} \mu_\text{B} \mathfrak{g}_{\text{e}} \sqrt{\frac{7N\mu_0\omega_\text{EPR}}{2V\hbar}}.
\end{align}

At $T > 0$, the optical response can be calculated using the transfer matrix method with the impedance $\sqrt{\mu_\mathrm{r}(\omega)/\epsilon_\text{r}} = 1/\sqrt{(1-\chi(\omega))\epsilon_\text{r}}$ given by the susceptibility $\chi(\omega)$ in Equation~\eqref{eq:chi} and the relative permittivity $\epsilon_\text{r} = n^2$.
By simulating the spectra, we can isolate the polariton peak position and fit the VRS to the experimental data by tuning $g_0$. The calculated VRS values are compared to the experimental values in Figure~\ref{Fig:Data2}\textbf{c} and \textbf{f}.


As we proved in Methods, the zero-temperature coupling strength $g_0$ is the one appearing in the Dicke model. The value of $g_0$ is theoretically estimated as follows. Because GGG is expressed as a simple cubic lattice with a lattice constant of $a = 1.238$\,nm and a unit cell includes 24 Gd$^{3+}$ ions, the density of magnetic dipoles is $N/V = 24/a^3$. Considering each magnetic dipole as a spin-7/2, we get
\begin{align}
    {g_0}/({2\pi})
    & = \sqrt{7.221\times \omega_\text{EPR}/(2\pi\; \mathrm{GHz})}\;\mathrm{GHz}
\end{align}
For $\omega_\text{EPR}/(2\pi) = 218\;\mathrm{GHz}$, we get $g_0/(2\pi) = 39.7\;\mathrm{GHz}$, which is close to the value of 47.5\;GHz obtained from fitting the data. 

%
The temperature dependence of the VRS (effective coupling strength) can be understood through the imaginary susceptibility $\mathrm{Im}[\chi(\omega)]$ (absorption in absence of cavity) shown in Figure~\ref{Fig:Theory}\textbf{c} ($T = 1.5$\,K) and \textbf{d} ($T = 250$\,K). At $T = 1.5$\,K, the probability $P_{m_\text{s}}$ is concentrated at the lowest level $m_\text{s} = -7/2$, as seen in Figure~\ref{Fig:Theory}\textbf{a}, and $\mathrm{Im}[\chi(\omega)]$ simply consists of a single Lorentzian involved with the lowest transition $-7/2$ to $-5/2$. At $T = 250$\,K, the population is distributed almost equally in all the eight levels, and $\mathrm{Im}[\chi(\omega)]$ consists of three Lorentzian functions (seven transitions) whose amplitude is much smaller than that at $T=1.5$\,K, reflecting the small population difference $P_{m_\text{s}-1}-P_{m_\text{s}} \ll 1$ in Equation~\eqref{eq:g2}.
The linewidth of these three peaks is sufficiently broad ($\gamma/(2\pi) = 80$\,GHz) and they are spectrally indistinguishable from a single broad Lorentzian with $\Gamma/(2\pi) = 120$\,GHz shown by dashed line in Figure \ref{Fig:Theory}\textbf{d}. 

\section*{Discussion}
As shown in Figure~\ref{Fig:Data2}\textbf{c} and \textbf{f}, our calculated VRS shows excellent agreement with the experimental data. Thus, we conclude that as the temperature is increased and more spin ensembles are thermally excited into higher-energy orientations, the interaction between the photons and spins is reduced until, at high temperatures, the system saturates. At low temperatures, the permeability is dominated by the lowest-energy Zeeman term, corresponding to the $m_\text{s} = -7/2 \rightarrow -5/2$ transition. Alternatively, around room temperature, each energy transition contributes nearly equally (see Figure~\ref{Fig:Theory}\textbf{c} and \textbf{d}), corresponding to a reduced coupling strength. This behavior marks a deviation from the boson-boson systems requisite to most light-matter coupling studies, which display no temperature dependence in the interaction strength other than a common thermal broadening process. 

The cavity system described here is, thus, a condensed matter system that can be accurately described by the Dicke Hamiltonian modeling of the interaction between a eight-level system and a photon bath. Here we examined the thermodynamic ($N\rightarrow \infty$) limit, where the Holstein--Primakoff transformation applies, and the Dicke Hamiltonian becomes the Hopfield Hamiltonian (with temperature-dependent coefficients). However, we find that while the system can be generally described by a boson-boson interaction picture, the light-matter coupling strength $g$ retains the fermionic character of the original Dicke model. Therefore, Zeeman-polaritons provide a unique opportunity to directly examine Dicke physics rather than simulating in analogous systems.

In conclusion, we studied the temperature and magnetic field dependence of EPR resonances in GGG and their coupling to FP cavity modes. The examination of spin-boson coupled systems is critical for further explorations of Dicke phenomena. Compared to other magnetic resonances, paramagnetic spins offer a unique avenue for exploring Dicke physics in condensed matter systems. Furthermore, because the coupling strength is a function of the population of each level in a multi-level system, the system is susceptible to thermal distributions and thus bears a dependence on the sample temperature. The ability to tune the coupling strength provides a distinct advantage of Zeeman polaritons for cavity-enabled technologies, nonlinear light-matter coupling, and advanced sensing. The versatile use of an array of cavity modes, when combined with the ability to tune $g$ with temperature, gives unprecedented control over the coupling strength.

\section*{Methods}
\subsection*{Terahertz Magnetospectroscopy}
We performed terahertz (THz) time-domain magnetospectroscopy~\cite{BaydinEtAl2021FO} in the Faraday geometry in transmission mode across two experimental setups. In the first one, THz pulses were generated using a (110) ZnTe crystal pumped by a Ti:sapphire amplifier (150\,fs, 1\,kHz, 0.8\,mJ, Clark-MXR, Inc., CPA-2010) via optical rectification. The part of the laser output was used as a probe beam with the time delay varied by a translation stage. The THz beam transmitted through the sample was probed by electro-optic sampling in another ZnTe crystal, resulting in a bandwidth of about 2.5\,THz. This THz setup was coupled to a helium-cooled superconducting magnet (Oxford Instruments Spectromag 10-T), with variable temperatures from 1.4\,K to 300\,K and static fields up to 10\,T. 

To probe Zeeman polaritons at higher magnetic fields, we utilized the Rice Advanced Magnet with Broadband Optics (RAMBO)~\cite{Noe2016OptExp,BaydinEtAl2021FO,Tay2022JPSJ}, a single-shot THz time-domain spectroscopy setup coupled with a pulsed magnet coil with fields up to 30\,T. Similarly to the first setup, a 775-nm optical pulse from an amplified Ti:sapphire laser (150\,fs, 1\,kHz, 0.8\,mJ, Clark-MXR, Inc., CPA-2001) was used to generate THz radiation via optical rectification in LiNbO$_3$ (limited THz bandwidth up to 1.6\,THz). The emitted THz radiation was focused on the sample, and then electro-optically sampled using ZnTe. Single-shot detection was accomplished using a reflective echelon and a fast CCD camera. The sample was placed at the center of the magnetic coil, coupled to a cryostat via a cold sapphire rod, allowing for temperatures down to 12\,K.

The THz electric field was measured as a function of time for all samples and in free space, which was used as a reference. Transmittance was calculated as the square of the ratio of the signal intensity to the reference for the Fourier-transformed THz time-domain signal.

\subsection*{Sample Preparation}
To form our cavity, we utilized (100)-cut single crystals of GGG purchased from MSE Supplies LLC. In order to adjust the FP cavity modes, formed by the GGG sample itself, the sample was mounted onto an Allied OltiPrep wafer polisher and thinned down with 400-grit silicon-carbide paper. 800- and 1200-grit silicon-carbide paper were used last to provide an even finish. After polishing, the sample was measured with an optical profilometer to ensure even thickness throughout the sample.

\subsection*{GGG Relative Permeability} \label{Sec:permeability}
We first consider the following Maxwell's equations 
\begin{subequations}
\begin{align}
\bm{\nabla}\times\bm{H} & = \frac{\partial\bm{D}}{\partial t}, \\
\bm{\nabla}\times\bm{E} & = - \frac{\partial\bm{B}}{\partial t},
\end{align}
\end{subequations}
where the electric displacement field $\bm{D} = \epsilon_0\epsilon_\mathrm{r}\bm{E}$ is connected to the electric field $\bm{E}$ via the vacuum and relative permittivities $\epsilon_0\epsilon_\mathrm{r}$, while the magnetic flux density $\bm{B} = \mu_0\mu_\mathrm{r} \bm{H}$ is expressed through vacuum and relative permeabilities $\mu_0 \mu_\text{r}$ as well as the magnetic field $\bm{H}$.
From the above Maxwell's equations, the dispersion relation of the Zeeman polaritons is expressed as
\begin{align} \label{eq:disp}
    \frac{c^2 k^2}{\omega^2} = \epsilon_\mathrm{r} \mu_\mathrm{r}(\omega).
\end{align}
To find an expression for the relative permeability $\mu_\text{r} (\omega)$, we derive the magnetic susceptibility $\chi(\omega)$ from linear response theory~\cite{Kubo1957-nz}.
To this end, we model the EPR absorption by the interaction Hamiltonian
\begin{align} \label{eq:Zeeman2}
    \hat{\mathcal{H}}_{\text{I}xy} = -\hat{\bm{d}} \cdot \bm{B}
\end{align}
describing the coupling of the magnetic moment $\bm{\hat{d}}=-\mathfrak{g}_{\text{e}} \mu_\text{B} \hat{\bm{s}}/ \hbar$ of a spin-$7/2$ particle to an AC magnetic flux density $\bm{B}$, while the unperturbed Hamiltonian $\hat{\mathcal{H}}_0$ is the one giving the energy levels \cite{Dai1988PRD} in Figure \ref{Fig:ExpScheme}\textbf{e} including the Zeeman interaction with the DC magnetic field $B^{\text{DC}}$, which is assumed to point in the positive $z$-direction.
Since we are interested in the susceptibility of the system due to an AC field only, we choose a direction perpendicular to the DC field, in particular the $x$-direction.
The magnetic susceptibility is then defined as the linear response of the magnetization $M_x  = N \braket{\hat{d}_x} / V$ in the $x$-direction, induced by $N$ independent magnetic dipoles in a volume $V$, i.e., 
\begin{align}
M_x(\omega) = \chi(\omega) B_x(\omega)/\mu_0
\end{align}
with magnetic susceptibility
\begin{align} \label{eq:chi_B}
    \chi(\omega)
    & = - \frac{N\mu_0}{V} \int_0^{\infty}\diff t\ \frac{\mathrm{e}^{i\omega t}}{i\hbar}
    \braket{[ \hat{d}_x(t), \hat{d}_x]}.
\end{align}
Using a thermal distribution $\hat{\rho} = \text{exp} \{ -\hat{\mathcal{H}}_0/(k_\text{B}T)\}/Z$ with partition function $Z$ between the Zeeman levels to evaluate the expectation value in Equation \eqref{eq:chi_B}, we find Equation \eqref{eq:chi}.
Thus, the susceptibility resolves to a sum of Lorentzian absorptions over each energy transition, where $s=7/2$ for GGG. Each energy $E_{m_\text{s}}$ is occupied with a probability 
\begin{align} \label{eq:probability}
    P_{m_\text{s}} = \frac{1}{Z} \exp\left(\frac{-E_{m_\text{s}}}{k_{\text{B}} T}\right)
\end{align}
appearing in the temperature-dependent coupling strength $g_{m_\text{s}}$ in Equation \eqref{eq:g2}.
Note that $g_0$ in Equation \eqref{Eq:g0}
is associated with the coupling strength of the system's ground state ($P_{-7/2}=1$), and thus, $\omega_\text{EPR}$ corresponds to the transition frequency between the two lowest Zeeman levels.
The matter damping rate $\gamma$ in Equation \eqref{eq:chi} has been introduced in the usual way as a consequence of regularizing the divergent integral in Equation \eqref{eq:chi_B}.

From the definition of the magnetic flux density $\bm{B} = \mu_0 (\bm{H} + \bm{M})$ and the relative permeability $\mu_\mathrm{r}(\omega)$ connecting the magnetic field and magnetic flux density $\bm{B} = \mu_0\mu_\mathrm{r}(\omega) \bm{H}$, we arrive at
\begin{align} \label{eq:mu_rB}
    \mu_\mathrm{r}(\omega) = \frac{1}{1-\chi(\omega)}.
\end{align}

\subsection*{Zeeman Interaction Coupling Strength}
We determine the light--matter coupling strength directly from the Zeeman interaction Hamiltonian, Equation~\eqref{eq:Zeeman}, for a collection of $N$ identical spins. The magnetic flux density is expressed with the photonic annihilation operator $\hat{a}_{x}$ with the $x$ polarization as
\begin{align} \label{eq:B-a}
    \hat{B}_x & = \sqrt{\frac{\hbar\mu_0\omega_{\text{cav}}}{2}} \frac{\hat{a}_x^{\dagger} + \hat{a}_x}{\sqrt{V}},
\end{align}
where we assume a spatially uniform magnetic flux density for simplicity. Here, we also assume that the $N$ spins are distributed in that cavity mode as
\begin{align} \label{eq:Zeeman3}
    \hat{\mathcal{H}}_{\text{I}N} =  \frac{\mathfrak{g}_{\text{e}} \mu_\text{B}}{\hbar} \hat{B}_x \sum_{i=1}^N\hat{s}_{i,x}.
\end{align}
Next, we approximate each spin as a two-level system consisting of a lower state $\ket{g} = \ket{\frac{7}{2},-\frac{7}{2}}$ and an upper state $\ket{e} = \ket{\frac{7}{2},-\frac{5}{2}}$. The spin operator in the Zeeman interaction, Equation \eqref{eq:Zeeman3}, is approximated as
\begin{align}
    \hat{s}_{i,x} \approx \Braket{\frac{7}{2},-\frac{5}{2}|\hat{s}_{i,x}|\frac{7}{2},-\frac{7}{2}}\hat{\sigma}_+ + \mathrm{H.c.} = \frac{\sqrt{7}}{2} \hbar \hat{\sigma}_{i,x}
\end{align}
Thus, the coupling Hamiltonian is approximated as
\begin{align}
    \hat{\mathcal{H}}_{\text{I}N} \approx  \frac{\sqrt{7}}{2} \mathfrak{g}_{\text{e}} \mu_\text{B} \hat{B}_x \sum_{i=1}^N\hat{\sigma}_{i,x}
    = \sqrt{7} \frac{\mathfrak{g}_{\text{e}} \mu_\text{B}}{\hbar} \hat{B}_x \hat{S}_x,
\end{align}
where $\hat{S}_x$ is a spin-$\frac{N}{2}$ operator.
Substituting Equation \eqref{eq:B-a} into this, we get
\begin{align}
    \hat{\mathcal{H}}_{\text{I}N} \approx  \frac{2g_0}{\sqrt{N}} \hat{S}_x (\hat{a}_x^{\dagger} + \hat{a}_x),
\end{align}
where we can find the coupling strength equivalent to $g_0$ in Equation \eqref{Eq:g0} for $\omega_{\text{cav}} = \omega_{\text{EPR}}$.

\section*{Acknowledgements}
T.E.K., J.D., H.X., A.B., and J.K. acknowledge support from the U.S. Army Research Office (through Award No. W911NF2110157), the Gordon and Betty Moore Foundation (through Grant No. A23-0150-001), the Robert A. Welch Foundation (through Grant No. C-1509), and the W. M. Keck Foundation (through Award No. 995764). T.E.K and J.K. acknowledge support from the Global Institute for Materials Research Tohoku and International Collaborations with Institute for Materials Research, Tohoku University. M.B. acknowledges support from the Japan Society for the Promotion of Science (through Grants No. JPJSJRP20221202 and No. JP24K21526).

\section*{Author Contributions}
J.K. and A.B. conceived and supervised the project. T.E.K. conceived experimental plans, performed THz measurements, analyzed all experimental data, simulated and fitted theoretical data, and prepared the manuscript under the supervision of J.K. and A.B.. J.D. and H.X. prepared samples. M.B. developed the theoretical model to describe the temperature dependence, and supervised T.A. and S.Y. in theoretical analysis. H.Y. and A.N. provided theory support for the magnetic response of GGG. M.B. and T.A. wrote the theory section in the manuscript. I.K. and J.T. conceived the single-shot detection technique. H.N. developed the 30-T pulsed magnet. All authors reviewed the manuscript. 

\section*{Competing Interests}
The authors declare no competing interests.

\bibliography{kritzell}


\end{document}